\documentclass[10pt]{paper}
\usepackage[margin=1in]{geometry}
\usepackage{graphicx}
\usepackage{authblk}
\usepackage{fancyhdr}
\usepackage{color,xcolor}
\usepackage[round,authoryear,semicolon]{natbib}
\usepackage{hyperref}
\usepackage{amssymb}
\usepackage{amsmath}
\usepackage{caption}
\usepackage{url}
\hypersetup{pdfpagelayout=OneColumn,colorlinks = true,urlcolor   = blue,citecolor  = blue,linkcolor  = blue}
\usepackage{adjustbox}
\usepackage{algpseudocode}
\usepackage[]{algorithm}
\usepackage{footnote}
\usepackage{bm}
\usepackage{setspace}
\usepackage{soul}
\usepackage{lineno}
\usepackage{parskip}
\title{Site characterization at downhole arrays by joint inversion of dispersion data and acceleration time series}
\author{Elnaz Seylabi\thanks{Department of Civil and Environmental Engineering, University of Nevada, Reno, NV 89557 (\url{elnaze@unr.edu})}, Andrew Stuart\thanks{Department of Computing and Mathematical Sciences, California Institute of Technology, Pasadena, CA 91125}, Domniki Asimaki\thanks{Department of Mechanical and Civil Engineering, California Institute of Technology, Pasadena, CA 91125}}
\date{}
\begin{document}
{\small \maketitle}
\begin{abstract}
\noindent
We present a sequential data assimilation algorithm based on the ensemble Kalman inversion to estimate the near-surface shear wave velocity profile and damping when heterogeneous data and a priori information that can be represented in forms of (physical) equality and inequality constraints in the inverse problem are available. Although non-invasive methods, such as surface wave testing, are efficient and cost effective methods for inferring Vs profile, one should acknowledge that site characterization using inverse analyses can yield erroneous results associated with the inverse problem non-uniqueness. One viable solution to alleviate the inverse problem ill-posedness is to enrich the prior knowledge and/or the data space with complementary observations. In the case of non-invasive methods, the pertinent data are the dispersion curve of surface waves, typically resolved by means of active source methods at high frequencies and passive methods at low frequencies. To improve the inverse problem well-posedness, horizontal to vertical spectral ratio (HVSR) data are commonly used jointly with the dispersion data in the inversion. In this paper, we show that the joint inversion of dispersion and strong motion downhole array data can  also reduce the margins of uncertainty in the Vs profile estimation. This is because acceleration time-series recorded at downhole arrays include both body and surface waves and therefore can enrich the observational data space in the inverse problem setting. We also show how the proposed algorithm can be modified to systematically incorporate physical constraints that further enhance its well-posedness. We use both synthetic and real data to examine the performance of the proposed framework in estimation of Vs profile and damping at the Garner Valley downhole array, and compare them against the Vs estimations in previous studies.
\end{abstract}
\section{Introduction}
Downhole arrays have been extensively used as testbeds by engineers and earth scientists for validation and improvement of predictive models of site response and physics-based ground motions. Strong motions recorded at depth, in particular, are widely sought after boundary conditions for the validation of one-dimensional wave propagation codes, as they minimize the uncertainty associated with source and path effects when studying problems of site response. To best serve as validation testbeds for equivalent linear and nonlinear site response analyses, however, downhole array sites should also be accompanied by reliable estimates of soil profiles--such as small strain shear modulus (or shear wave velocity), damping and nonlinear soil properties--as well as their variability. 

Site characterization efforts at strong motion stations include shear wave velocity measurements at multiple locations, using both invasive (e.g. PS-logging) and/or non-invasive methods (e.g. spectral analysis \citep{stokoe1994characterization} and multi-channel analysis of surface waves \citep{foti2000multistation}). Nonlinear soil properties, on the other hand, are most frequently estimated from empirical correlations of published laboratory data \citep[e.g.,][]{darendeli2001development}, or, in rare cases, are measured from undisturbed samples retrieved at the site. Laboratory measured properties, however, are not always reliable estimates of in-situ soil properties, an incompatibility associated with sample disturbance, measurement error and inherent field-scale spatial variability of soil properties. Downhole array ground motion records have been used instead by several researchers to infer in-situ soil parameters. More specifically, low amplitude motions have been used to constrain the Vs profile and damping \citep[e.g.,][]{assimaki2006attenuation} while high amplitude motions have been used to parameterize nonlinear soil behavior in moderate to large strains \citep[e.g.,][]{assimaki2011wavelet,chandra2015situ,seylabi2018bayes}.

While it is commonly assumed that invasive methods are more reliable than non-invasive methods to retrieve Vs profiles, recent studies have highlighted that the latter are more efficient and cost effective, while yielding uncertainties comparable to the order of invasive methods \citep{garofalo2016interpacific2,teague2016site}. Still, one should acknowledge that site characterization using inverse analyses, that is surface wave testing or downhole array data, can yield erroneous results associated with the inverse problem non-uniqueness. Furthermore, recent studies have shown that significantly different Vs profiles that satisfy the error thresholds of the inversion process, can still result in similar linear viscoelastic seismic site response and amplification factors \citep[e.g.,][]{foti2009non}.

One viable solution to alleviate the inverse problem ill-posedness is to enrich the prior information and/or the data space with complementary data.
In the case of non-invasive methods, the pertinent data {are} the dispersion curve of surface waves, typically resolved by means of active source methods at high frequencies and passive methods at low frequencies. 
Horizontal to vertical spectral ratios (HVSR) have also been used extensively to approximate a site's predominant frequency, and therefore constrain the velocity structure at depth. Furthermore, the joint inversion of HVSR with dispersion data has been used successfully to improve the site characterization accuracy \citep[e.g.,][]{arai2005s,pina2016inversion,molnar2018application,lunedei2015review}. 

At strong motion arrays, recorded acceleration time-series, which include both body and surface waves, can also be used to complement dispersion data. Because of the complementary characteristics of the body and surface waves -- the former carrying information in the form of travel time and the latter in the form of near-surface dispersive characteristics -- we anticipate that formulating the inversion problem using both dispersion data and acceleration time-series should reduce the margins of uncertainty in the Vs profile estimation. 

Inversion methods that are used in site characterization include the stochastic direct search method, which is based on the neighborhood algorithm \citep[e.g.,][]{wathelet2004surface}; the uniform Monte Carlo method \citep[e.g.,][]{socco2008improved}, which finds an ensemble of models minimizing the data misfit; and the fully Bayesian Markov Chain Monte Carlo method \citep[e.g.,][]{molnar2010bayesian}, which provides the most probable and quantitative uncertainty estimates of the Vs profile. In this paper, we present a framework based on the ensemble Kalman inversion and use it to examine whether and how the joint inversion of dispersion and downhole array data improves site characterization estimates. We also show how one can systematically incorporate a priori information in terms of physical constraints in the ensemble Kalman inversion to improve the problem well-posedness. 

In the rest of this paper, we first explain the ingredients of the framework in \S Inverse Problem. In \S Site Characterization--Synthetic Data and \S Site Characterization--Real Data, we perform numerical experiments using synthetic and real data simulated or recorded at the Garner Valley Downhole Array (GVDA) site -- one of the best-studied and best-instrumented sites in Southern California. More specifically, in \S Site Characterization--Synthetic Data we use synthetic data to study how the combined data sets improve the Vs profile estimation relative to the individual data.  Next,
in \S Site Characterization--Real Data, we use recorded acceleration time-series by the array and experimental dispersion data to invert for the Vs profile and damping, and compare our results to the inverted Vs profiles from previous studies. Finally, we provide concluding remarks in \S Conclusion.

\section{Inverse Problem}
\subsection{Problem Formulation}
\noindent We consider the problem of finding $u$ from a series of data sets $y_i$ where
\begin{equation}
y_i = G_i(u)+\eta_i\,.
\label{eq:inv-prob}
\end{equation}
$u \in \mathbb{R}^k$ consists of $k$ unknown/uncertain parameters, $y_i \in \mathbb{R}^{m_i}$ consists of $m_i$ observations spanning the i$^{th}$ data space, $\eta_i\in\mathbb{R}^{m_i}$ is the noise represented as independent zero-mean Gaussian noise with covariance $\Gamma_i$; and  
$G_i$ is a nonlinear function (referred to as the forward model) that maps the parameter space to the i$^{th}$ data set. In this paper, we work with two data sets, the dispersion curve that depicts discrete phase velocity values of surface waves as a function of frequency, and the discrete acceleration time series recorded by the array instruments at different depths. If we combine these two data sets, the inverse problem of \eqref{eq:inv-prob} is modified as follows:
\begin{equation}
\begin{bmatrix}
y_1\\y_2
\end{bmatrix}
=
\begin{bmatrix}
G_1(u)\\
G_2(u)
\end{bmatrix}
+
\begin{bmatrix}
\eta_1\\\eta_2
\end{bmatrix}
\rightarrow y = G(u)+\eta, \quad 
\Gamma = \begin{bmatrix}
\Gamma_1 & 0 \\ 0 & \Gamma_2
\end{bmatrix}\,.
\end{equation}
We define the covariance matrix of the Gaussian noise as follows:
\begin{align}
\Gamma_1 = \left[\beta_1\text{diag}(\max|y_1|\bm{1})\right]^2, \quad \Gamma_2 = \left[\beta_2\text{diag}( y_2)\right]^2
\label{eq:gamma}
\end{align}
where $\beta_1$ and $\beta_2$ determine the noise levels for $y_1$ and $y_2$ which are acceleration time series and dispersion data, respectively.

For the two data sets relevant to the problem in hand, the forward models are described below: 
\begin{itemize}
\item[i)] For the theoretical dispersion curve we use the transfer matrix approach originally developed by \citet{thomson1950transmission} and \citet{haskell1953dispersion} and later modified by \citet{dunkin1965computation} and \citet{knopoff1964matrix}. This approach requires the solution of an eigenvalue problem, for which we use the well known software GEOPSY \citep{wathelet2005array}. We should mention here that all the examples considered in this study are normally dispersive, so results presented here only require the dispersion curve for the first mode of Rayleigh waves. However, the framework is general enough to allow for higher modes to be incorporated in the inversion process. 

\item[ii)] For the theoretical acceleration time series, we consider wave propagation in a horizontally stratified layered soil of total thickness $H$ and shear wave velocity $V_s(z)$ varying with depth $z$. Given an acceleration time series at $z = H$ (i.e., the borehole sensor depth), we compute the soil response numerically using a finite element model and we use the extended Rayleigh damping \citep{phillips2009damping} to capture the nearly frequency independent viscous damping $\xi$ in time domain analyses. We also assume that Vs below $z=H$ is constant, which corresponds to an elastic bedrock idealization. 
\end{itemize}

\subsection{Algorithm}
\noindent To solve the inverse problem involving the two data sets just described, alone or in conjunction, we use a sequential data assimilation method \citep{evensen2009data}  based on ensemble Kalman inversion \citep{iglesias2013ensemble}, a methodology pioneered in the oil reservoir community \citep{O1,O2}. In this algorithm, we first define an initial ensemble consisting of  $N$ particles. In its most basic form, the ensemble Kalman inversion can regularize ill-posed inverse problems through the subspace property where the solution found is in the linear span of the initial ensemble employed \citep{chada2019tikhonov}. Then, at each iteration $j$, we use the forward model predictions $G(u_j^{(n)})$ and the observation data $y_{j+1}$ to update these particles sequentially:
\begin{equation}
\label{eq:update}
u_{j+1}^{(n)} = u_j^{(n)} + C_{j+1}^{uw}(C_{j+1}^{ww}+\Gamma)^{-1}(y_{j+1}^{(n)}-G(u_j^{(n)})) \quad \text{for} \quad n = 1, \dots, N \,,
\end{equation}
where $y_{j+1}^{(n)}$ can be either identical to $y_{j+1}$ (the observation data) or found by adding to $y_{j+1}$ identical and independently distributed zero-mean Gaussian noise $\eta_{j+1}^{(n)}$ with distribution 
the same as that of $\eta$; the matrices $C_{j+1}^{uw}$ and $C_{j+1}^{ww}$ are empirical covariance matrices that can be computed at each iteration based on predictions and the ensemble mean $\bar{u}_{j+1}$ using the following equations.
\begin{equation}
C_{j+1}^{uw} = \frac{1}{N}\sum_{n=1}^{N}(u_j^{(n)}-\bar{u}_{j+1})\otimes(G(u_j^{(n)})-\bar{G}_j)
\end{equation}
\begin{equation}
C_{j+1}^{ww} = \frac{1}{N}\sum_{n=1}^N(G(u_j^{(n)})-\bar{G}_j)\otimes(G(u_j^{(n)})-\bar{G}_j)
\end{equation}
and
\begin{equation}
\bar{u}_{j+1} = \frac{1}{N}\sum_{n=1}^Nu_j^{(n)}, \quad \bar{G}_j = \frac{1}{N}\sum_{n=1}^{N}G(u_j^{n}) \,.
\label{eq:mean}
\end{equation}

To enforce physical and prior knowledge systematically, \citet{albers2019ensemble} provide an efficient procedure to impose constraints within the ensemble Kalman filtering framework; they use the solution of a constrained quadratic programming to update particles that violate the enforced constraints; in the absence of constraints, the optimization delivers the update formulae above. In the case of linear
equality and inequality constraints the approach is readily implemented using standard optimization algorithms. We briefly summarize the procedure in the case of linear inequality constraints:
\begin{equation}
Au\leq a
\end{equation}
and when the problem is formulated in the range of the covariance matrix $C_{j+1}$:
\begin{equation}
    C_{j+1} = \begin{bmatrix}
    C_{j+1}^{uu} & C_{j+1}^{uw}\\
    {C_{j+1}^{uw}}^T & C_{j+1}^{ww}
    \end{bmatrix}
\end{equation}
where
\begin{equation}
    C_{j+1}^{uu} = \frac{1}{N}\sum_{n=1}^{N}(u_j^{(n)}-\bar{u}_{j+1})\otimes(u_j^{(n)}-\bar{u}_{j+1})\,.
\end{equation}
It should be noted that the empirically computed covariance is the sum of rank one matrices and its rank is at most $N-1$. For this problem setting, $N-1$ is generally less than $k+m_1+m_2$ and therefore the covariance matrix is not invertible. To overcome this issue, \citet{albers2019ensemble} reformulate the problem in the range of the covariance in which they seek the solution as a linear combination of a given set of vectors. For more details see \citet{albers2019ensemble}. As such, for each violating particle we seek a vector $b^{(n)}$ that minimizes the cost function $J_{j,n}(b)$ defined as
\begin{equation}
\label{eq:min}
J_{j,n}(b) := \frac{1}{2}|y_{j+1}^{(n)}-G(u_j^{(n)}) - \frac{1}{N}\sum_{m=1}^N b_m (G(u_j^{(m)})-\bar{G}_j)|_\Gamma^2+\frac{1}{2N}\sum_{m=1}^N (b_m)^2
\end{equation}
and subject to
\begin{equation}
\label{eq:constrain}
ABb \leq a-A{u}_{j}^{(n)}
\end{equation}
where
\begin{equation}
Bb = \frac{1}{N} \sum_{m=1}^N b_m(u_j^{(m)}-\bar{u}_{j+1})\,.
\end{equation}
Next, we use the computed $b^{(n)}$ to update each violating particle $u_j^{(n)}$ as follows:
\begin{equation}
\label{eq:update2}
u_{j+1}^{(n)} = u_j^{(n)}+\frac{1}{N}\sum_{m=1}^{N}b_m^{(n)}(u_j^{(m)}-\bar{u}_{j+1})\,.
\end{equation} 
We have implemented this algorithm -- summarized in Algorithm \ref{alg} -- and have  verified its accuracy in the numerical results section of \citet{albers2019ensemble}. 

\begin{algorithm}[htbp]
\caption{Constrained ensemble Kalman inversion algorithm formulated in range of covariance}
\label{alg}
\begin{algorithmic}[1]
\State Choose $\{u_{0}^{(n)}\}^{N}_{n=1}$, $j=0$
\State Calculate forward model application $\{G(u_j^{(n)})\}^{N}_{n=1}$
\State Update $\{{u}_{j+1}^{(n)}\}^{N}_{n=1}$ from \eqref{eq:update}
\For {$n=1:N$}
\If{${u}_{j+1}^{(n)}$ violates constraints in \eqref{eq:constrain}}
\State{$b^{(n)} \leftarrow$ argmin of \eqref{eq:min} subject to \eqref{eq:constrain}}
\State{Update $\{{u}_{j+1}^{(n)}\}$ from (\ref{eq:update2})}
\EndIf
\EndFor
\State $j \gets j+1$, go to 2.
\end{algorithmic}
\label{alg:7}
\end{algorithm}

\section{Site Characterization: Synthetic Data}\label{sec:synthetic}
\begin{figure}[htbp]
\centering
{\includegraphics[trim = 2cm 16cm 2cm 2cm,clip,width = 0.9\textwidth]{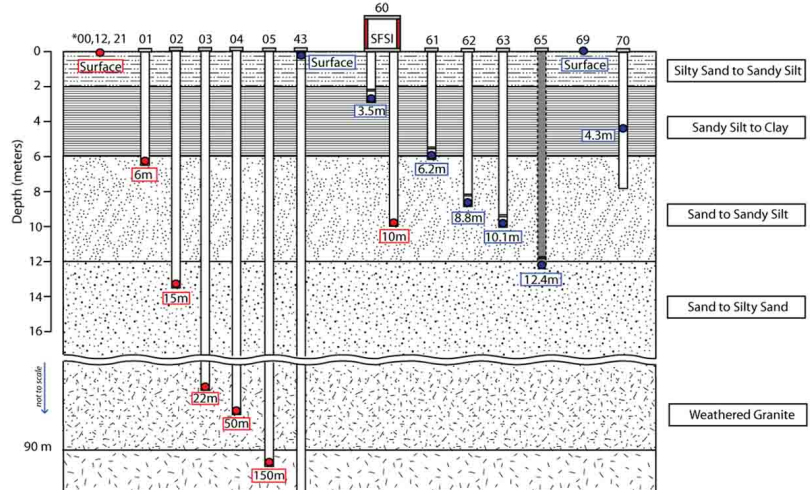}}
\caption{Garner Valley downhole array cross section (cf. \S Data \& Resources).}
\label{fig:GVDA}
\end{figure}

\noindent 
We first use synthetic data to evaluate the importance of joint inversion of downhole array and dispersion data in near-surface site characterization using the proposed framework. In this numerical experiment, we use the site conditions at the GVDA in Southern California, which is also the site where we test the algorithm using recorded ground motion data in the following sections of this work. GVDA is located in a narrow valley, and the near-surface structure consists of an ancestral lake bed with soft alluvium down to 18-25 m depth, overlaying a layer of weathered granite; the competent granitic bedrock interface is located at 87-m depth according to \citet{bonilla2002borehole} and  varies across the valley {\citep{teague2018measured}}. Figure \ref{fig:GVDA} shows the site geology and the layout of sensors at different depths including accelerometers (red boxes) and pressure transducers (blue boxes). Several invasive and non-invasive Vs measurements have been carried out at this site in the past; the most recent surface wave measurements were performed in October 2016 by \citet{teague2018measured}, who developed Vs profiles at each of the three surface accelerometer locations (i.e., at the location of sensors 00, 12, and 21 in Figure~\ref{fig:GVDA}). 

To generate \emph{ground truth} downhole array data, we use acceleration time series recorded at GVDA from an event with magnitude 4.28. We also consider the Vs profile shown in Table \ref{tab:vs} as the target profile; and we assume mass density $\rho=1800$ kg/m$^3$, Poisson's ratio $\nu=0.3$ and damping ratio $\xi=0.04$ for all layers. We next use these input parameters in the forward models described above to compute the acceleration time-history at the ground surface and the first mode Rayleigh wave dispersion curve; and use these simulated data as observations in the inversion process.
\begin{table}[htbp]
\centering
\caption{Vs profile at Garner Valley Downhole Array site used for synthetic data experiments; the values up to depth 100 m are from \citet{gibbs}.}
\begin{adjustbox}{max width = \textwidth}
\begin{tabular}{ccc}
\hline
Layer & Thickness [m] & Shear Wave Velocity (Vs) [m/s]\\
\hline
1 & 0.0--18.0  & 220\\
2 & 18.0--64.5 & 580\\
3 & 64.5--150 & 1300\\
4 & Half-space & 2600\\
\hline
\end{tabular}
\end{adjustbox}
\label{tab:vs}
\end{table}

When reliable invasive measurements are not available at the site, prior information on the thickness of the soil layers is also unavailable. To overcome this shortcoming, we use a fine discretization of the profile  shown in \eqref{eq:thick}, with increasing thickness increments $\Delta h$ with depth, ranging from $\Delta h = 5$ m to $\Delta h= 25$ m. This selection resulted in $r=15$ layers for the $H=150$ m thick profile (from the surface to the depth of borehole sensor 05 shown in Figure~\ref{fig:GVDA}). Our forward model also considers elastic bedrock boundary conditions beyond depth 150 m, which we characterize by a thin layer of thickness 1 m in \eqref{eq:thick}.
\begin{equation}
{\Delta h}~[\text{m}] =\{5,5,5,5,5,5,10,10,10,10,15,15,25,24,1\}
\label{eq:thick}
\end{equation}

Assuming that the Vs is constant in each layer, and the profile is normally dispersive, we enforce monotonic behavior by posing the following linear inequality constraints, and lower and upper bounds for the very first and last layers. We assume that the Vs profile can change monotonically between 50 m/s at surface and 5000 m/s at the bedrock, which is wide enough search space for our inversion. As we mentioned before, enforcing such constraints reduces the velocity model complexity and the inverse problem ill-posedness.
\begin{equation}
V_{s,i} \leq Vs_{s,i+1} \quad \text{for} \quad i = 1, \dots, r-1, \quad V_{s,1} \ge 50, \quad V_{s,r} \leq 5000\,.
\end{equation}
For estimation of the small-strain damping, we consider the range $0.001\leq \xi \leq 0.1$, while the density and Poisson's ratio for each layer are assumed constant and equal to the values we use in our forward model simulation: $\rho=1800$ kg/m$^3$, Poisson's ratio $\nu=0.3$.

\begin{figure}[htbp]
\centering
{\includegraphics[trim = 2.5cm 19cm 2.5cm 2.5cm,clip,width = \textwidth]{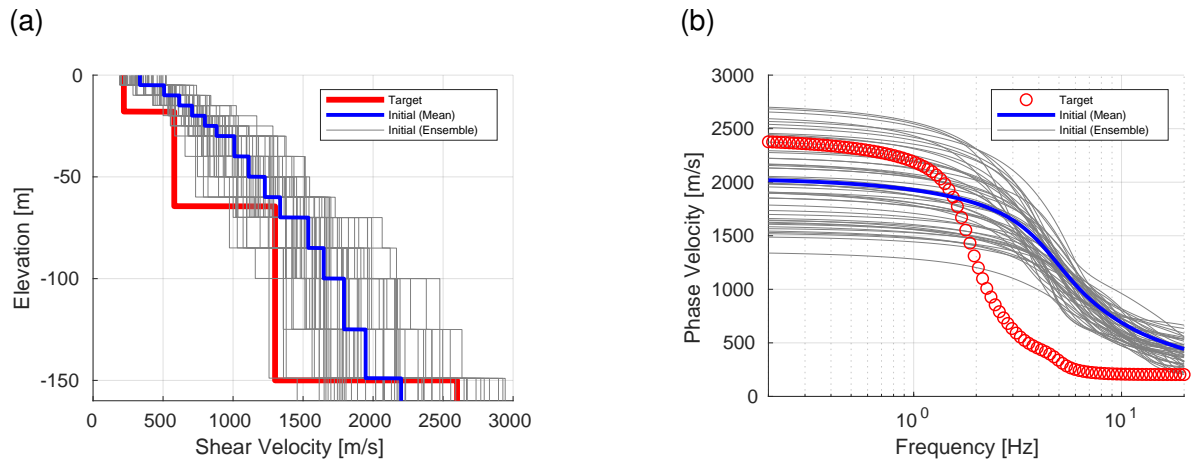}}
\caption{(a) Vs profiles of the initial ensemble along with the ensemble mean and the target profile -- the vertical scale is elevation relative to the ground surface; (b) Dispersion curves computed using the initial ensemble, ensemble mean an target Vs profiles.}
\label{fig:initial}
\end{figure}

We next initiate the algorithm by creating the initial ensemble of particles. We draw particles from the uniform distributions for both Vs and damping, and scale them appropriately as follows:
\begin{equation}
P(V_{s,i}) = \sqrt{\frac{z_i}{H}}\left [ 5 + 10 U(0,1) \right ], \quad P(\xi) = 0.1 + 0.3U(0,1)
\label{eq:dist}
\end{equation}
where $U(0,1)$ is the uniform distribution between 0 and 1 and $z_i$ is the depth of the bottom of layer $i$. Then we use the projection in \eqref{eq:project} to enforce constraints on each particle $u^{(n)}$ that violates the constraints in the initial ensemble.
\begin{equation}
u_0^{(n)} = \underset{u}{\text{argmin}}~\frac{1}{2}|u-u^{(n)}|^2 \quad \text{subject to} \quad Au\leq a
\label{eq:project}
\end{equation}
where $u^{(n)}$ is before we enforce the constraints, and $u_0^{(n)}$ is its projected counterpart that satisfies the constraints. Figure~\ref{fig:initial}a shows $N=50$ realizations of the projected Vs profiles drawn from the above distribution along with the ensemble mean and the target profile. Figure~\ref{fig:initial}b shows the corresponding dispersion curves for each particle, the mean and the target profile. In the next subsections, we use the ensemble Kalman inversion algorithm explained in \S Inverse Problem to estimate the Vs and/or damping profiles of a horizontally layered soil. We specifically consider three test cases to generate the data space $y$ for the inversion. These include: only dispersion data (both a complete and an incomplete set), only downhole array data, and fusion of downhole and dispersion data. We only estimate damping for cases that involve downhole array data.

\subsection{Dispersion Data As Inversion Data Space}
\noindent Here, we assume that both active and passive surface wave testing results are available; the former is usually used to resolve dispersion data at higher frequencies (shallow layers) and the latter to constrain dispersion data at long periods (deep layers). 
The size of the testing arrays and lateral variability in the depth of the bedrock across these arrays may cause difficulties in resolving the dispersion curve continuously for a wide range of frequencies. To examine how data completeness affects the uncertainty of the inverted profile at strong motion arrays, we define two data sets: a complete and incomplete dispersion curve. In the incomplete data set, dispersion data corresponding to frequencies smaller than 0.3 Hz, and frequencies between 0.45 and 2.35 Hz, are missing. 

\begin{figure}[htbp]
{\includegraphics[trim = 2.5cm 19cm 2.5cm 2.5cm,clip,width = \textwidth]{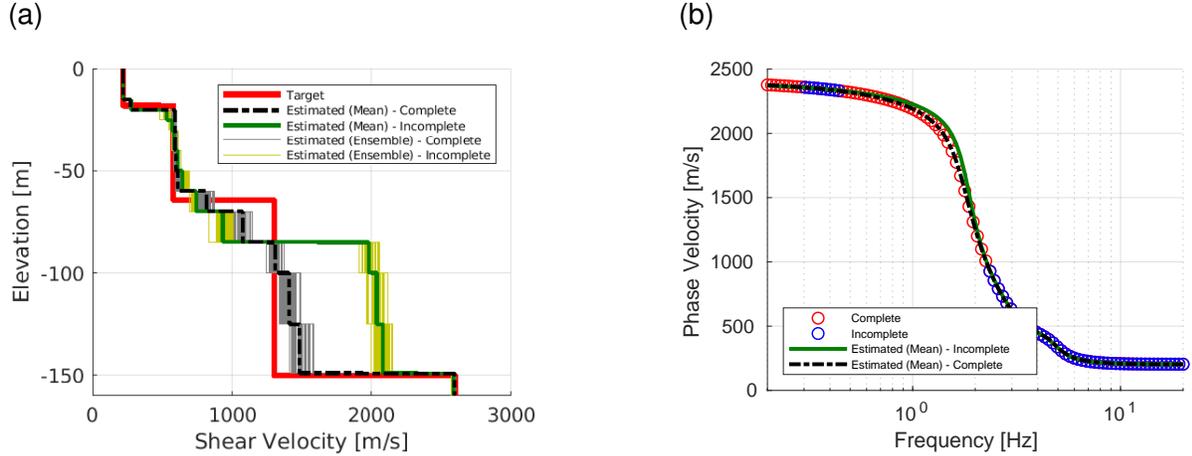}}
\caption{(a) Final estimated ensembles along with the ensembles' mean and target Vs profile -- the vertical scale is elevation relative to the ground surface; (b) Dispersion curves computed using the ensemble mean along with the complete and incomplete datasets used in the inverse analysis.}
\label{fig:estimate}
\end{figure}

We next use Algorithm \ref{alg} to iteratively update the particles. Figure~\ref{fig:estimate}a shows the estimated Vs profiles, which are computed from the ensemble mean in the last iteration, using the complete and incomplete data sets, and Figure~\ref{fig:estimate}b shows the computed dispersion curves compared against the data sets used as observations. We should mention here that the stopping criterion for all performed inverse analyses is reaching 100 iterations; this number is more than enough to make the mean of the ensemble stabilized around the reported final estimates. The solution non-uniqueness is evident in Figure~\ref{fig:estimate}: the dispersion curves of both profiles are in excellent agreement with the curve associated with the target profile, whereas the two inverted profiles show significant differences. Unsurprisingly, for the case of simulated dispersion data, slight deviation is only observed for the dispersion data of the incomplete set, exactly in the frequency range where information is missing.

\subsection{Downhole Array Ground Motion Recordings As Inversion Data Space}
\noindent In this section, we repeat the numerical experiment described above (i.e. estimation of Vs and damping) using only downhole array ground motion recordings. The forward problem comprises of propagating the ``recorded'' (known) motion at depth $z=H$ to the ground surface, and minimizing the misfit between the ground surface motion forward predictions and the surface acceleration time series. Downhole arrays are usually instrumented sparsely (see Figure~\ref{fig:GVDA} for example), which may contribute to the solution non-uniqueness and uncertainty. To reflect this issue, we only use the instrument at 150~m depth as input and the ground surface motion as output, which is the most common configuration of a downhole array (e.g. of the Japanese strong motion network, KiK-Net). 

\begin{figure}[htbp]
\centering
{\includegraphics[trim = 2.5cm 13.5cm 2.5cm 2.5cm,clip,width = \textwidth]{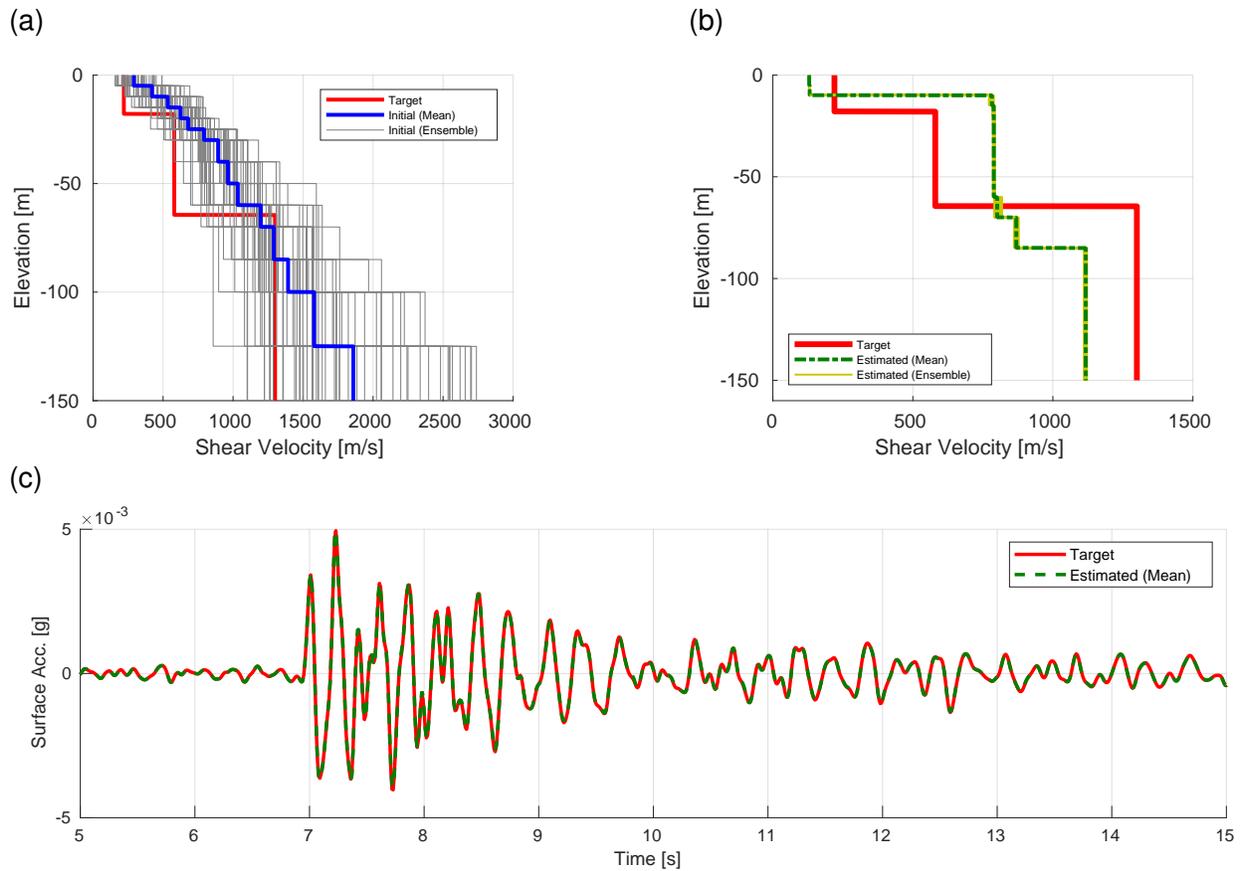}}
\caption{(a) Initial ensemble of particles along with the ensemble mean and target profile -- the vertical scale is elevation relative to the ground surface; (b) Estimated ensemble of particles along with the ensemble mean and target profile; (c) Computed surface acceleration time series using the target Vs profile and damping as well the mean of the estimated ensemble.
}
\label{fig:estimateDH}
\end{figure}

To compare with the dispersion data inversion, we first drew the initial particles from the same distribution as in \eqref{eq:dist} but the algorithm was not successful in finding the profile that can reproduce the output data. However, when we slightly shifted the initial ensemble to Vs values closer to the target solution (see \eqref{eq:distDH} and Figure~\ref{fig:estimateDH}a), the algorithm successfully converged to a Vs profile and damping ratio that captures the ground surface acceleration. We should mention here that in this case, where the downhole recorded motion is used as prescribed boundary condition, our forward model is appropriately adjusted to a layered soil on rigid bedrock, in which the thickness of the last layer is 25 m.
\begin{equation}
P(V_{s,i}) = \sqrt{\frac{z_i}{H}}\left [ 4 + 10 U(0,1) \right ], \quad P(\xi) = 0.1 + 0.3U(0,1)\,.
\label{eq:distDH}
\end{equation}
Figure~\ref{fig:estimateDH}b shows the estimated Vs profile after 100 iterations and Figure~\ref{fig:estimateDH}c shows the computed surface accelerations using this profile. The final estimate for damping is $\xi = 0.0412$ which is close to the target damping ratio of 0.04 used to generate the synthetic data. While we see clear differences between the inverted and the target Vs profiles, both predict identical ground surface motion when subjected to the borehole strong motion record, which once again demonstrates the solution non-uniqueness when the data space is sparse.

\subsection{Dispersion And Downhole Array Data As Joint-Inversion Data Space}
\noindent So far we have tried using the two heterogeneous data sets separately: first, we used the dispersion data that provided constraints on the dispersive characteristics of surface waves; and then we used downhole array data, which provided constraints on the travel time of body (shear) waves through the soil layers. As shown in the last two examples, the inversion algorithm could not recover the target profile in either case whereas all forward models captured the target observations exceptionally well. 

\begin{figure}[!htbp]
\centering
{\includegraphics[trim = 2.5cm 19cm 2.5cm 2.5cm,clip,width = \textwidth]{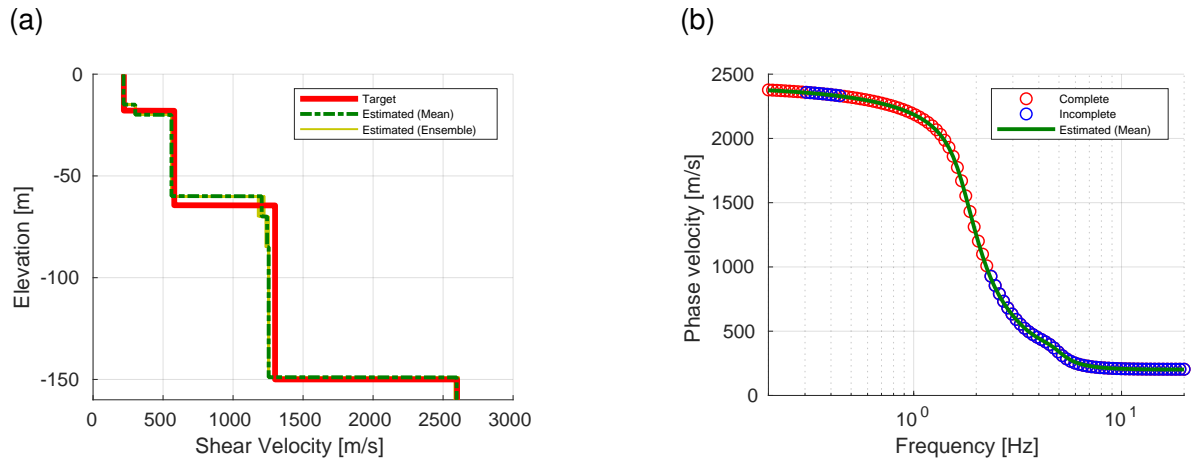}}
\caption{(a) Final estimated ensemble along with the ensemble mean and target profile -- the vertical scale is elevation relative to the ground surface; (b) Dispersion curve using the estimated ensemble mean along with both complete and incomplete datasets used in the inverse analysis.
}
\label{fig:estimatefuse}
\end{figure}

In this section, we study the effects of fusing these two complementary data sets. We create the initial ensemble from \eqref{eq:dist}, and consider $y$ as the combination of the incomplete dispersion data and surface acceleration time series. We use $\beta_1=0.01$ and $\beta_2=0.01$ to define the Gaussian noise covariance matrix. Using the same inversion algorithm, Figure~\ref{fig:estimatefuse} shows the estimated Vs profile and the computed dispersion results. As can be readily seen, by combining the two data sets, we are able to recover the Vs profile with depth; as expected, the profile matches both the incomplete dispersion curve and the acceleration time series on ground surface. The latter is almost identical to the time series shown in Figure~\ref{fig:estimateDH}c and therefore is not repeated here. Furthermore, the algorithm recovers a constant damping ratio of $\xi=0.0413$; recall that our synthetic example uses a constant damping $\xi=0.04$ for all layers. This example shows the effectiveness of joint inversion to successfully recover the target Vs profile and damping ratio among possible solutions that can fit the observations well individually in the inverse problem setting. In the next subsection, we test the algorithm effectiveness for more complex profiles and noise-contaminated synthetic data before proceeding to an example using measured dispersion data and recorded downhole array ground motions.

\subsection{Profile Complexity And Noise Effects On Inverted Vs Profiles}\label{sec:complex}
\noindent As shown in the previous subsection, joint inversion of downhole array and dispersion data helps improve the estimation of the Vs profile. So far, however, we have only used noise-free synthetic data. Prior to introducing field recorded data, we test the performance of the framework using more complex profiles and noise contaminated synthetic data. It should be noted that the complex profile we will use in the next example is not intended to be representative of the Vs profile at GVDA; rather, we generate it to assess the framework's capability in dealing with more complex cases.

\subsubsection{Profile Complexity Effects}

\begin{figure}[!htbp]
\centering
{\includegraphics[trim = 2.5cm 13.5cm 2.5cm 2.5cm,clip,width = \textwidth]{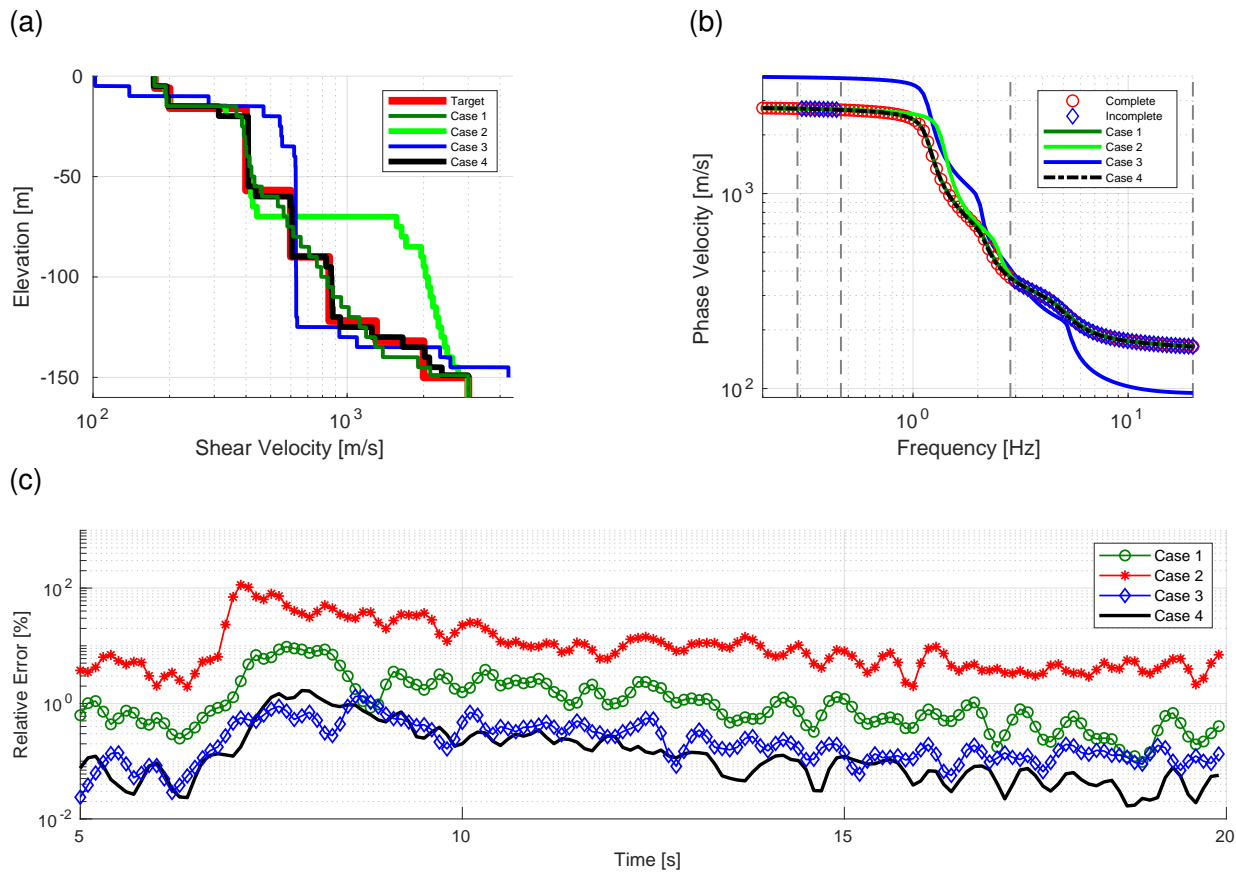}}
\caption{(a) Finale estimate of the ensembles' mean along with the target Vs profile -- the vertical scale is elevation relative to the ground surface; (b) Dispersion curves computed using the ensembles' mean along with the complete and incomplete datasets used in the inverse analysis; (c) Smoothed relative error between the surface acceleration dataset and those computed using the estimated ensembles' mean.
}
\label{fig:complex}
\end{figure}

\noindent
Here, we use the same framework to estimate the Vs profile for a more complex case. To capture the complex Vs, we use a more refined discretization of the profile, i.e., $\Delta h = 5$ m, namely 30 unknown Vs parameters. We again consider four types of data sets: 
\begin{itemize}
    \item complete dispersion data (Case 1);
    \item incomplete dispersion data where phase velocity values for frequencies $f\leq 0.3$ Hz and $0.46\leq f\leq 2.83$ Hz are missing (Case 2);
    \item downhole array data only at the surface $z=0$ (Case 3);
    \item fusing data sets in Case 2 and Case 3 (Case 4).
\end{itemize}
For Case 1 and Case 2 we estimate 31 parameters including the elastic bedrock Vs and the Vs from zero to $H$. For Case 3, we estimate 31 parameters including the Vs from zero to $H$ and damping, and for Case 4, we estimate 32 parameters including the Vs from zero to $H$, elastic bedrock Vs, and damping. Figure \ref{fig:complex}a and  \ref{fig:complex}b show the estimated Vs profiles after 100 iterations along with the resulting dispersion curves. On the other hand, Figure~\ref{fig:complex}c shows the smoothed relative error of the surface acceleration computed using different estimated profiles. It should be noted that the absolute error for Cases 3 and 4 is very small. Again, we notice that joint inversion of dispersion and downhole array data can significantly improve the estimated profile. Furthermore, increasing the number of soil layers does not affect the performance of the algorithm presented here and it can correctly resolve the impedance contrasts in the target Vs profile.

\vspace{10pt}
\subsubsection{Noise Effects}

\begin{figure}[htbp]
\centering
{\includegraphics[trim = 2.5cm 13.5cm 2.5cm 2.5cm,clip,width = \textwidth]{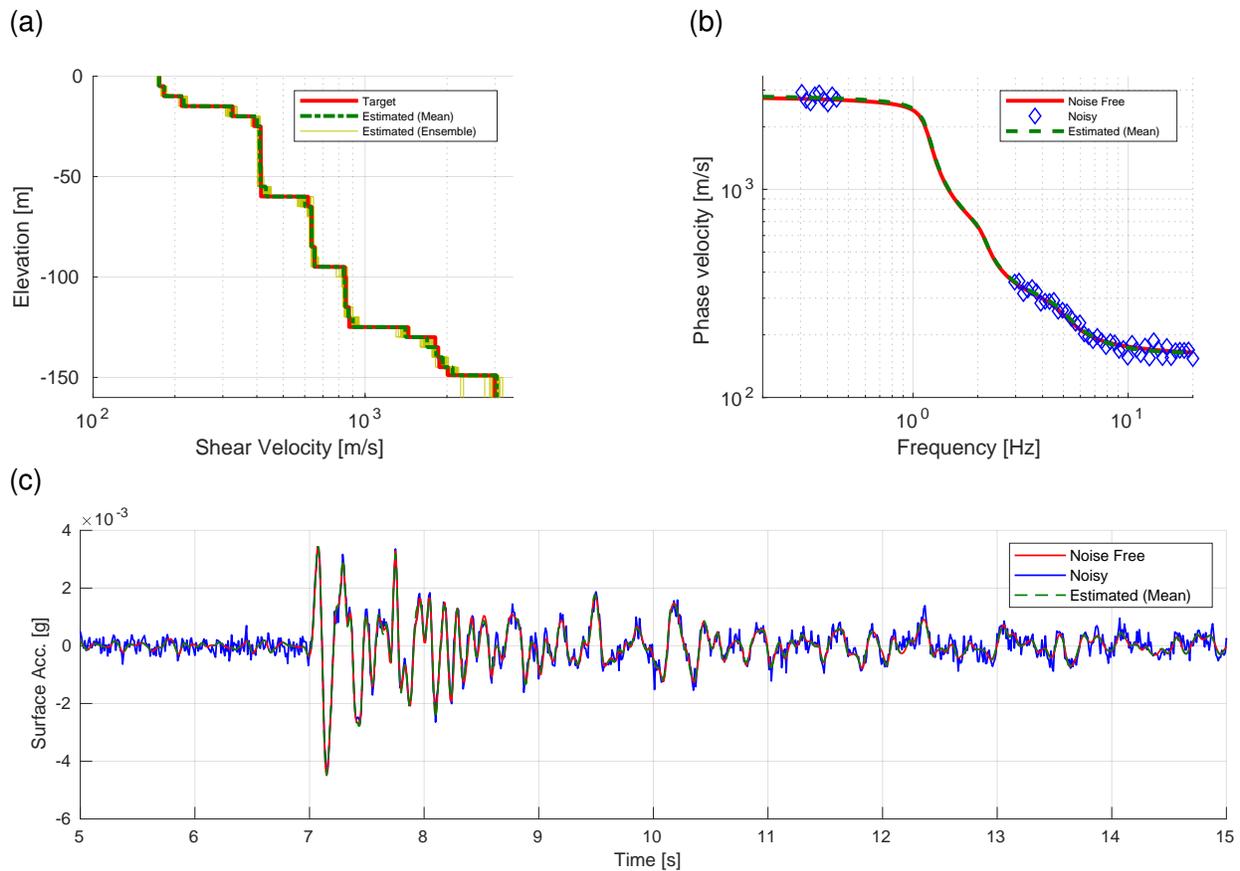}}
\caption{
(a) Final estimated ensemble along with the ensemble man and target Vs profile -- the vertical scale is elevation relative to the ground surface; (b) Dispersion curve computed using the estimated ensemble mean along with the noise free and incomplete noisy dispersion data used in the inverse analysis; (c) Surface acceleration time series computed using the estimated ensemble mean along with the noise free and noise contaminated acceleration data used in the inverse analysis.
}
\label{fig:noise}
\end{figure}

\noindent
To assess the framework robustness when the recorded data is noisy, we next consider Case 4 where we use the combined data sets to estimate Vs profile and damping. For generating the noise contaminated data sets $y_1$ and $y_2$, we add a zero mean Gaussian noise where we use \eqref{eq:gamma} to define the covariance matrix considering $\beta_1=0.05$ and $\beta_2=0.05$. Then, within the ensemble Kalman inversion iterations, we consider $\beta_1=0.025$ and $\beta_2 = 0.025$ to define $\Gamma$.
Figure~\ref{fig:noise}a shows the estimated Vs profile, and Figures~\ref{fig:noise}b and \ref{fig:noise}c show the theoretically computed dispersion curve and surface acceleration time series compared to the data (both clean and noisy); this exercise again shows the successful recovering of the Vs profile and damping in presence of the noisy data.

\section{Site Characterization: Real Data}\label{sec:gvda}

\begin{figure}[!htbp]
\centering
{\includegraphics[trim = 2.5cm 20.5cm 2.5cm 2.5cm,clip,width = \textwidth]{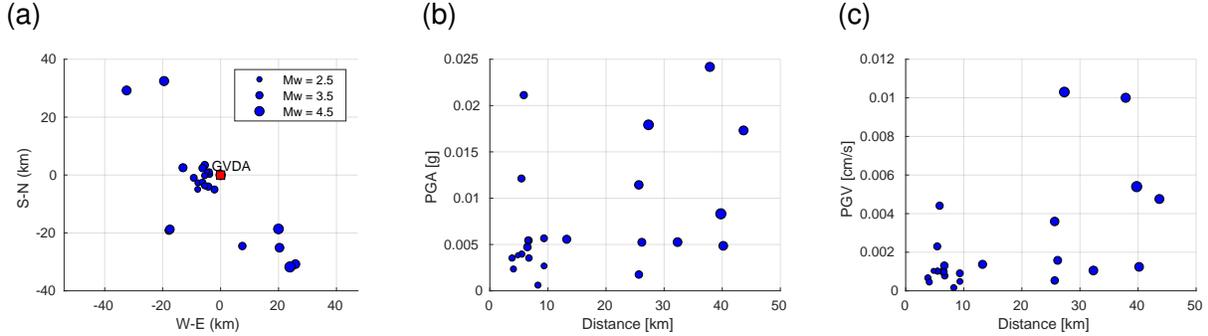}}
\caption{(a) Epicentral distance, (b) PGA and (c) PGV of the events considered for Vs estimation at GVDA.}
\label{fig:gv}
\end{figure}

\noindent In this section, we use the same inversion algorithm to estimate the Vs profile and damping at GVDA  using field dispersion data and recorded acceleration time series. For the downhole array data, we use a curated dataset (cf. \S Data \& Resources), comprising 30 events recorded from 2006 to 2016 with ground surface peak ground acceleration (PGA) greater than 10 gal. The events are recorded by vertical and aligned horizontal accelerometers from 501 m depth to the surface. From these records, we here focus on those with magnitude Mw~$\leq$~5 to minimize the likelihood of nonlinear response contaminating the recorded ground motions. In total we consider 23 events in what follows. Figure~\ref{fig:gv} shows the epicenter of the events along with the PGA and peak ground velocity (PGV) as a function of the epicentral distance.

As dispersion data set, we use the mean field measurements associated with the north accelerometer location 00 by \citet{teague2018measured}, who performed both active-source multi-channel analysis of surface waves and passive source microtremor array measurement testing at GVDA. Using the downhole array data $a_e = (a_{00}^e, a_{01}^e, a_{02}^e, a_{03}^e, a_{04}^e)$ for each event $e$ and dispersion data $V_r$, the observation can be formed as $y_e=(a_e, V_r)$ for $e=1,\dots, 23$. For each event, we use the inversion algorithm to estimate both $V_{s,e}$ and damping $\xi_e$. Please note that $a_{00}$, $a_{01}$, $a_{02}$, $a_{03}$, $a_{04}$, $a_{05}$ are accelerations recorded at $z=0, 6, 15, 22, 50$ and 150 meters respectively (see Figure~\ref{fig:GVDA}). All acceleration time series were filtered using second order bandpass Butterworth filter with frequency range [0.1, 10] Hz. 

\begin{figure}[htbp]
\centering
{\includegraphics[trim = 2.5cm 16cm 2.5cm 2.5cm,clip,width = \textwidth]{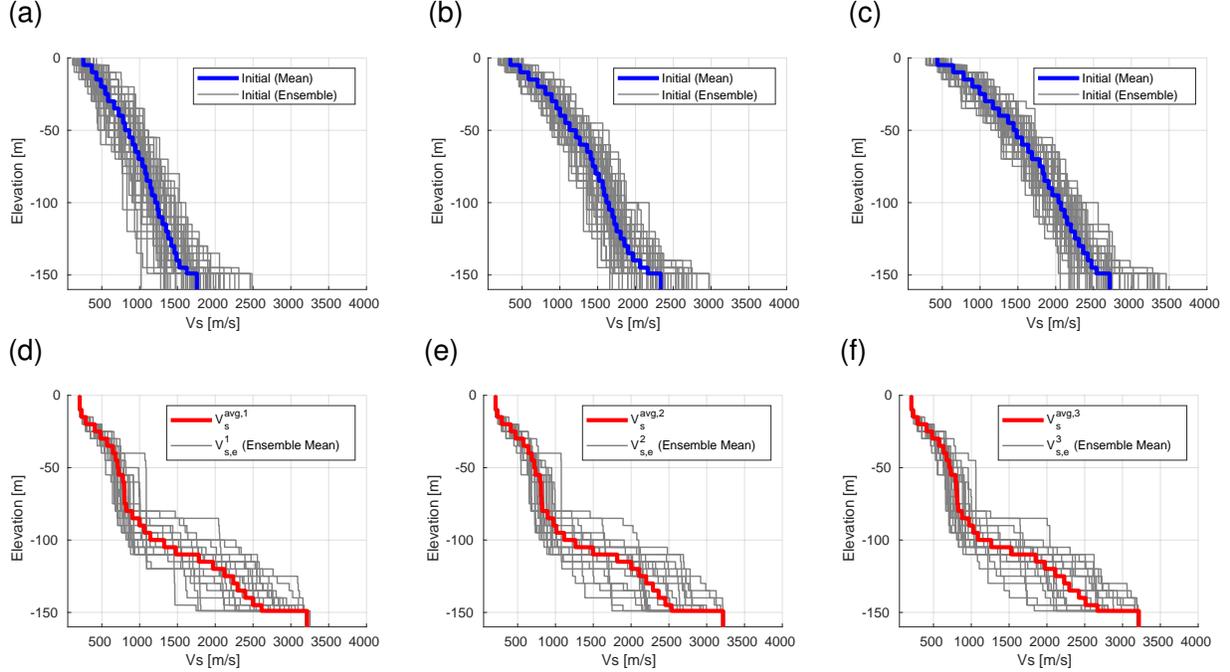}}
\caption{Initial ensemble of particles along with the ensemble mean: (a) $i=1$, (b) $i=2$, (c) $i=3$; Final estimated ensemble mean for events $e=1,\dots,23$ along with the average Vs profile: (d) $i=1$, (e) $i=2$, (f) $i=3$. The vertical scale is elevation relative to the ground surface.}
\label{fig:vsreal}
\end{figure}

\begin{figure}[htbp]
\centering
{\includegraphics[trim = 2.5cm 20.5cm 2.5cm 2.5cm,clip,width = \textwidth]{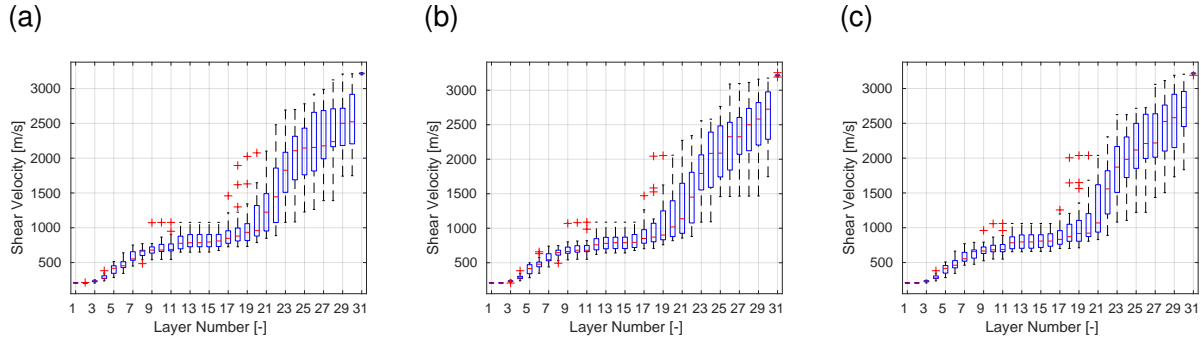}}
\caption{Box plot of the estimated Vs at each layer for the considered a priori distributions: (a) $i=1$, (b) $i=2$, (c) $i=3$. The central mark indicates the median, and the bottom and top edges of the box indicate the 25th and 75th percentiles, respectively. The whiskers extend to the most extreme data points not considered outliers, and the outliers are plotted individually using the `+' symbol.}
\label{fig:vsuq}
\end{figure}

\begin{figure}[htbp]
\centering
{\includegraphics[trim = 2.5cm 20.5cm 2.5cm 2.5cm,clip,width = \textwidth]{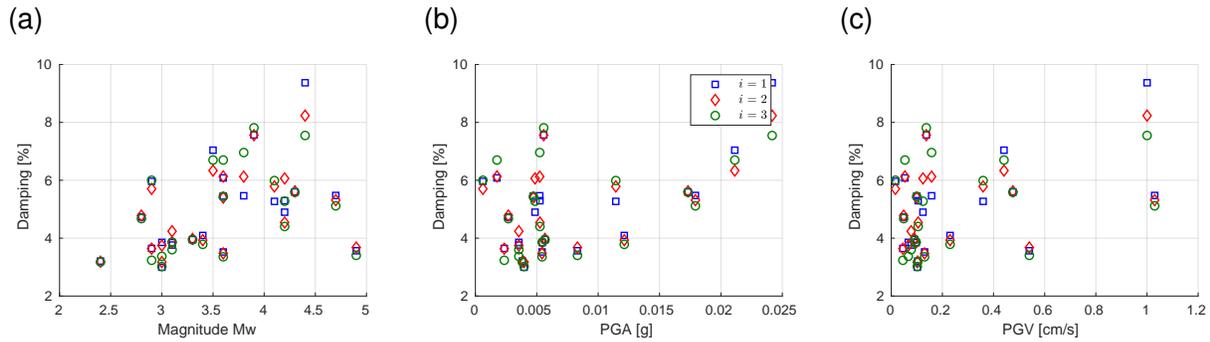}}
\caption{Final estimated ensemble mean of damping ratio for events $e=1,\dots, 23$ and as function of (a) event magnitude, (b) event PGA, (c) event PGV.}
\label{fig:damp}
\end{figure}

\begin{figure}[!htbp]
\centering
{\includegraphics[trim = 2.5cm 20.5cm 2.5cm 2.5cm,clip,width = \textwidth]{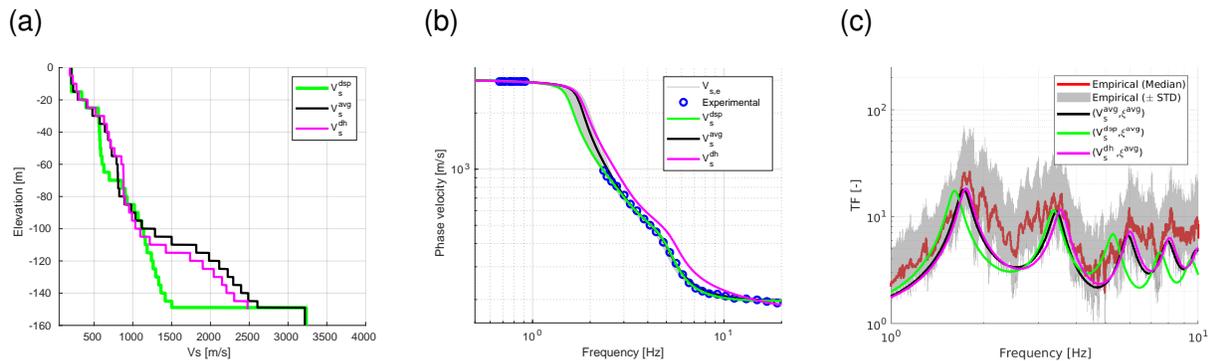}}
\caption{
(a) Estimated average Vs profile -- the vertical scale is elevation relative to the ground surface; (b) Dispersion curves computed using different average Vs profiles compared against the experimentally available dispersion data used in the inverse analysis; (c) Theoretical transfer functions computed using different average Vs profiles compared against the median of empirical transfer functions and their standard deviation.
}
\label{fig:vr}
\end{figure}

To start the inversion, we consider three initial ensembles with 50 particles each (see Figures~\ref{fig:vsreal}a, \ref{fig:vsreal}b, and \ref{fig:vsreal}c); and we seek to estimate the Vs and damping for $23\times 3 = 69$ cases. Doing so will allow us to assess the effects of using different prior distributions on the resulting profiles. For profile discretization, we consider uniform layers with thickness of 5~m. {It is worth mentioning that in what follows we will also study the effects of considering thicker layers for profile discretization.} Figures~\ref{fig:vsreal}d, \ref{fig:vsreal}e and \ref{fig:vsreal}f show the final estimated ensemble means for each event $e$ and each initial ensemble $i$ along with the average Vs profile $V_s^{\text{avg},i}$. 
\begin{equation}
V_{s}^{\text{avg},i} = \frac{1}{23}\sum_{e=1}^{23} V_{s,e}^i, \quad i = 1, 2 , 3\,.
\end{equation}
With the exception of small discrepancies at depths larger than 100~m, the profiles recovered by considering each prior and the corresponding ensemble are almost identical. Figure~\ref{fig:vsuq} shows the box plot of the estimated Vs at each layer, one for each of the prior distributions that we used to select our particle ensembles. Again, the trend in recovered Vs and associated uncertainty is nearly identical for all three prior Vs distributions. The resulting damping ratios (i.e., the ensemble mean for each $e$ and $i$) are shown in Figure~\ref{fig:damp} as a function of the event magnitude Mw, PGA and PGV;  the average damping is $\xi^\text{avg} = 0.049$ over 69 events and it appears that higher damping estimation is associated with higher ground motion intensities, which is very likely the manifestation of nonlinear response.
To evaluate the effectiveness of joint inversion, we next use the second initial ensemble ($i=2$) to estimate the Vs profile using only the dispersion data and only the downhole array data. Figure~\ref{fig:vr}a shows the estimated Vs profiles $V_s^{\text{dsp}}$ and $V_s^{\text{dh}}$ for dispersion and downhole array data, respectively; $V_s^\text{avg}$ is the average over all 69 events that we used to search the fused data space. As shown, $V_s^\text{dh}$ closely follows the $V_s^{\text{avg}}$ while $V_s^{\text{dsp}}$ starts deviating after $z$ around 30~m. Figures \ref{fig:vr}b and \ref{fig:vr}c show the effects of these inverted Vs profiles on the site signatures, i.e., dispersion curves and site transfer functions (TFs). The empirical TF for each event is obtained by dividing the Fourier transform of the recorded acceleration signal on the surface $a_{00}^{e,\dagger}$ by that of sensor $a_{05}^{e,\dagger}$ at depth $z=H=150$~m. Figure~\ref{fig:vr}c shows the median TF and its standard deviation among the considered 23 events. It is interesting to note that:
\begin{enumerate}
    \item although $V_s^{\text{dh}}$ is following the same trend as $V_s^{\text{avg}}$, the resulting dispersion curve cannot capture the experimental data while they capture the joint inversion TF quite well.
    \item while $V_s^{\text{dsp}}$ captures the dispersion data, the associated TF is different from those cases that downhole array data is incorporated in the inversion.
    \item Joint inversion of the dispersion and downhole acceleration data changes the inverted dispersion data only in the frequency range where phase velocity data are not available. 
    \item While peaks in the theoretical TFs are well-aligned with the empirical TF, it is worth noting that deviations at higher modes are possibly due to the three dimensional effects.
\end{enumerate}

\begin{figure}[!htbp]
\centering
{\includegraphics[trim = 2.5cm 20.5cm 2.5cm 2.5cm,clip,width = \textwidth]{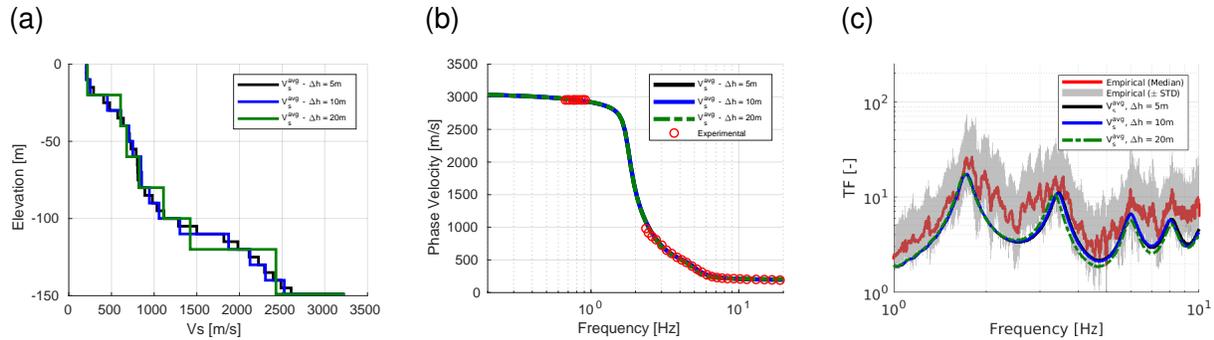}}
\caption{
(a) Estimated average Vs profile -- the vertical scale is elevation relative to the ground surface; (b) Dispersion curves computed using different average Vs profiles compared against the experimentally available dispersion data used in the inverse analysis; (c) Theoretical transfer functions computed using different average Vs profiles compared against the median of empirical transfer functions and their standard deviation.
}
\label{fig:coarse}
\end{figure}

To assess the effects of the inverse problem parameterization on inverted Vs profiles, we consider two more cases of discretizing the soil height with $\Delta h = 10$ m and $\Delta h = 20$ m and repeating the joint inversion. Figure~\hbox{\ref{fig:coarse}} shows the estimated Vs profiles and associated dispersion curves and transfer functions. These results suggest that decreasing the layer thickness, i.e., increasing the number of Vs values to be estimated, does not necessarily have adverse effects on the well-posedness of the inverse problem in hand. Furthermore, our previous numerical experiments suggest that fine discretization is successful to capture impedance contrasts at least when dealing with synthetic data. That being said, it is worth noting that, in many cases, it may not be clear which parameterization produces a ``better'' answer than others.  Therefore, to fully understand the uncertainties,  it may be necessary to consider multiple parameterizations.  Also, it is worth noting that our future research is aimed at addressing this issue systematically by estimating both layer thicknesses and Vs values.

\begin{figure}[!htbp]
\centering
{\includegraphics[trim = 2.5cm 8.5cm 2.5cm 2.5cm,clip,width = \textwidth]{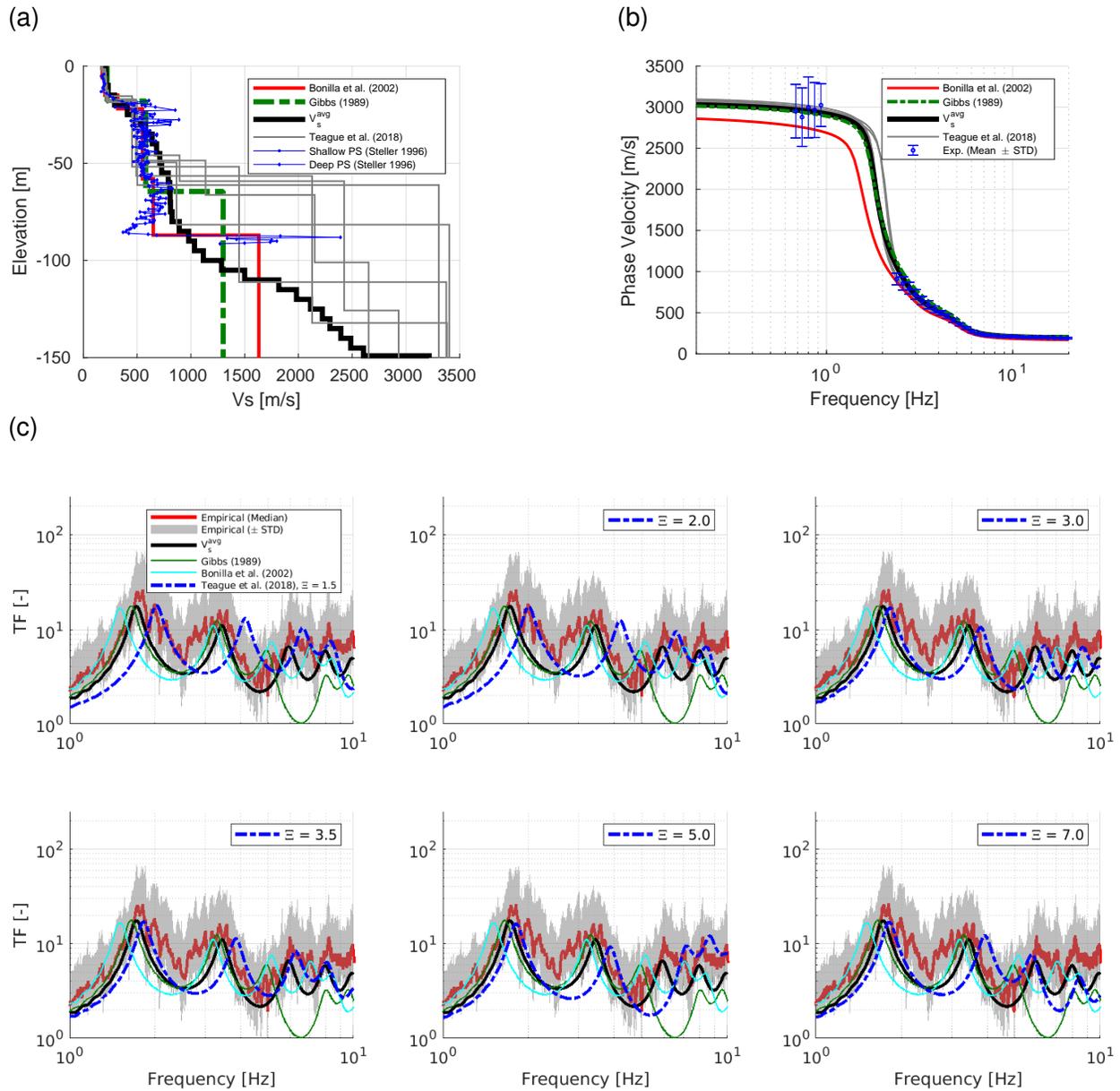}}
\caption{Comparison of (a) Vs profiles (the vertical scale is elevation relative to the ground surface), (b) dispersion curves and (c) transfer functions at GVDA.}
\label{fig:comp}
\end{figure}

Site characterization of Garner Valley has also been the subject of several geotechnical and geophysical studies, and therefore we compare our results to those available in open literature. More specifically, shallow and deep PS suspension logging results were provided by \citet{stellar1996new}. \citet{gibbs} used a three component geophone in a 100 m deep borehole, and determined the P and S wave velocities using the conventional methods of travel-time plots and straight line segments. On the other hand, \citet{bonilla2002borehole} used a trial-and-error approach to find the best Vs profile that minimizes the misfit between the synthetic and real downhole accelerometer time series for weak motions; they used the velocity models by \citet{gibbs,pecker1993garner} as the initial guess. Finally, \citet{teague2018measured} used the active and passive surface wave measurements to invert for the Vs profile using the dispersion data. They also used the HVSR curves to further constrain the inversion results by comparing the first mode frequency of theoretical TFs obtained from the inverted Vs profiles against the mean first mode frequency obtained from the experimental HVSR curves. These results along with those computed in this study are shown in Figure~\ref{fig:comp} in terms of Vs profile, theoretical dispersion curves, and TFs.
In all cases except for \citet{teague2018measured} we assume that $\nu = 0.45$ and $\rho = 1800$~kg/m$^3$. For \citet{teague2018measured} results, we use the data provided by the first author to compute the dispersion curves and TFs considering layering ratio $\Xi$ ranging from 1.5 to 7. Furthermore, for all computed TFs, we assume that the damping ratio is equal to the estimated average damping ratio, i.e, $\xi^{\textrm{avg}}=0.049$. Again, and as correctly identified by {\citet{teague2018measured}}, it is evident that lack of dispersion data in the moderate frequency range has resulted in discrepancies among the interpreted Vs profiles. Moreover, as shown in Figure~\ref{fig:comp}c, TFs of coarser profiles with $\Xi = 3, 3.5, 5, 7$ are in better agreements with those associated with this study.

\section{Conclusion}\label{sec:conclude}
\noindent In this paper, we introduced a sequential data assimilation approach based on ensemble Kalman inversion for near-surface site characterization. Our method was based on heterogeneous data set fusion and assimilation with prior knowledge, which we introduced in terms of inequality constraints to the Vs and/or small strain damping. To characterize the general trend of the Vs profile, we used the piece-wise constant function with a known number of layers. Through a series of synthetic experiments, we demonstrated the inverse problem solution non-uniqueness when dispersion data or acceleration time series were used in isolation and showed how the joint inversion of these complementary data could improve the Vs estimation. We also showed that increasing the number of layers can help capture more complex profiles without affecting the performance of the algorithm. Lastly, we tested the algorithm on real data using the Garner Valley Downhole Array site as our testbed, and compared the inverted Vs profile against previous site characterization studies. Our study showed that inversion uncertainties, such as the ones described by \citet{teague2018measured}, may be attributed to incomplete dispersion data in the medium frequency range. Future non-invasive testing that will help complete the available dispersion data across the entire frequency range of interest will help refine inverse algorithms such as the one presented here.

\section{Data and Resources}
\noindent 
The GVDA downhole array data are available at \url{ http://nees.ucsb.edu/curated-datasets} and Figure \ref{fig:GVDA} is downloaded from \url{http://nees.ucsb.edu/facilities/GVDA} (last accessed March 2020). The theoretical dispersion curves are computed using GEOPSY software package \texttt{gpdc} installed from \url{http://www.geopsy.org/download.php?platform=src&branch=testing&release=3.1.1}. Python code for performing ensemble Kalman inversion with constraints will be released at the github account of the first author (\url{https://github.com/elnaz-esmaeilzadeh}).

\bibliographystyle{apalike}
\bibliography{reference}

\end{document}